\documentclass[%
 reprint,
 amsmath,amssymb,
 aps,
]{revtex4-2}

\usepackage{mathptmx}  % Times font for serif
\usepackage{helvet}    % Helvetica font for sans-serif
\usepackage{courier}   % Courier font for monospaced

\usepackage{graphicx}% Include figure files
\usepackage{dcolumn}% Align table columns on decimal point
\usepackage{bm}% bold math
\usepackage{xcolor}

\begin{document}

\preprint{APS/123-QED}

\title{Enhanced \textit{Q} factor and robustness of photonic bound states in the continuum\\ merging at locally bent trajectories}% Force line breaks with \\

\author{Huayu Bai}
\email{huayu.bai@aalto.fi}
\author{Andriy Shevchenko}
 \email{andriy.shevchenko@aalto.fi}
\author{Radoslaw Kolkowski}
 \email{radoslaw.kolkowski@aalto.fi}
\affiliation{%
 Department of Applied Physics, Aalto University, P.O.Box 13500, Aalto FI-00076, Finland
}%

\begin{abstract}
Bound states in the continuum (BICs) in planar photonic structures have attracted broad scientific interest owing to their exceptional capability to confine light. Topological robustness of certain BICs allows them to be moved in the momentum space by tuning the geometric parameters of the structure. In this work, we study such a BIC in a one-dimensional periodic grating, and find that its momentum-space position can be made a non-monotonic function of a geometric parameter, forming a locally bent ``V''-shaped trajectory. We show that, near the turning point of this trajectory, the robustness of the BIC and its $Q$ factor can be greatly enhanced. We tune such ``V-BICs'' to almost merge with a symmetry-protected BIC at the $\Gamma$-point. This creates a ``K''-shaped ultrahigh-$Q$ region containing a BIC with a much higher and more stable $Q$ factor compared to the ordinary merging BICs. The ``K-BICs'' are also found to provide a strong enhancement of the $Q$ factor in finite gratings over an extremely wide range of geometric parameters. Our findings enable further advancements in the development of ultrahigh-$Q$ BICs and their applications.
\end{abstract}
\maketitle
%\tableofcontents

Achieving efficient light confinement and field enhancement has long been one of the central goals of research in nanophotonics~\cite{review-FKshrink}. Optical bound states in the continuum (BIC) have been especially promising in that context. They are a type of resonances that have their radiation loss eliminated by destructive interference in the far-field, and thus they can theoretically have an unlimited $Q$ factor~\cite{observationbic,review-azzam,review-hsu,review-kang}. In recent years, micro- and nanostructures supporting BICs or quasi-BICs with ultrahigh $Q$ factors have been demonstrated on various platforms, including photonics crystals~\cite{merging,bic-mini,supercriticalcoupling}, metasurfaces~\cite{metasurface-RK,NL-highq-metasurface} and compact nanostructures~\cite{miebic1,wgmbic1-laser}. Diverse applications and optical phenomena have been realized using BICs, including lasing~\cite{laser-bic-2017,laser-mergingbic,laser-minibic,laser2-minibic,merging-pre-laser,wgmbic1-laser}, sensing~\cite{sensing-index-biomolecular,sensing-molecular,sensing-refractiveindex,sensing-refractiveindex2,sensing-thickness}, nonlinear frequency conversion~\cite{NL,NL-highq-metasurface,NL-doubleresonant-phc,NL-merging-metasurface}, as well as controlling the polarization and enhancing the emission of light~\cite{chiralbic1-metasurface,chiralbic2-metasurface,chiralbic3-metasurface,polarizationcontrol-metasurface,ugr}.

The most common BICs can be classified into symmetry-protected BICs and accidental BICs~\cite{review-hsu}. Symmetry-protected BICs are typically located at the $\Gamma$ point in the momentum space, while accidental BICs usually have a non-zero in-plane momentum. In the momentum space, both types of BICs are located at the centers of the far-field polarization vortices, being characterized by integer topological charges~\cite{toponature,experimenttopovortex}. The conservation of global topological charge makes BICs robust against small geometry variations and allows some of them to be moved along a continuous trajectory in the momentum space. This enables the construction of a merged BIC~\cite{merging} by moving multiple BICs into a single point in the momentum space. Merging BICs show an enhanced $Q$ factor across a wide range of in-plane momentum, boosting the $Q$-versus-$k$ scaling law from $Q\sim1/k^2$ to $Q\sim1/k^6$. This makes them more robust to the out-of-plane scattering~\cite{merging,bic-mini}. Therefore, merging BICs are considered as a promising platform for a variety of applications based on enhanced light-matter interactions~\cite{merging-pre-laser,laser-minibic,laser-mergingbic,NL-merging-metasurface,NL-douleresonant-merging,merging-NL-faltband}.

However, the full potential of merging BICs is still unexplored. Further improvements can be achieved, e.g., by merging BICs with higher topological charges~\cite{merginghighorder,merging-highorder-2-two-accidental,jumping-bic-on-degenerate-band}, which has been shown to boost the scaling law to $Q\sim1/k^8$. It has also been discovered that merging BICs can be constructed without the symmetry-protected BICs~\cite{mergingoffgamma}, thus allowing them to appear at arbitrary points in the momentum space, and boosting the $Q$ factor at large incidence angles. Recently, we have proposed an asymmetry-compensation method that makes it possible to realize merging BICs in planar structures without up-down mirror symmetry~\cite{symmetrycompensation,bai24_}. Last but not least, several works have attempted to increase the robustness of merging BICs by coupling two metasurfaces to each other~\cite{merging-parameterspace-two-metasurface} or by merging BICs in a geometric parameter space~\cite{merging-parameterspace-bandfolding,merging-parameterspace-plasmonicparticle}. Despite these advances, there is still room for achieving merging BICs with even higher $Q$ factors and better stability.
\begin{figure}[hbt!]
    \centering
    \includegraphics[width=1\linewidth]{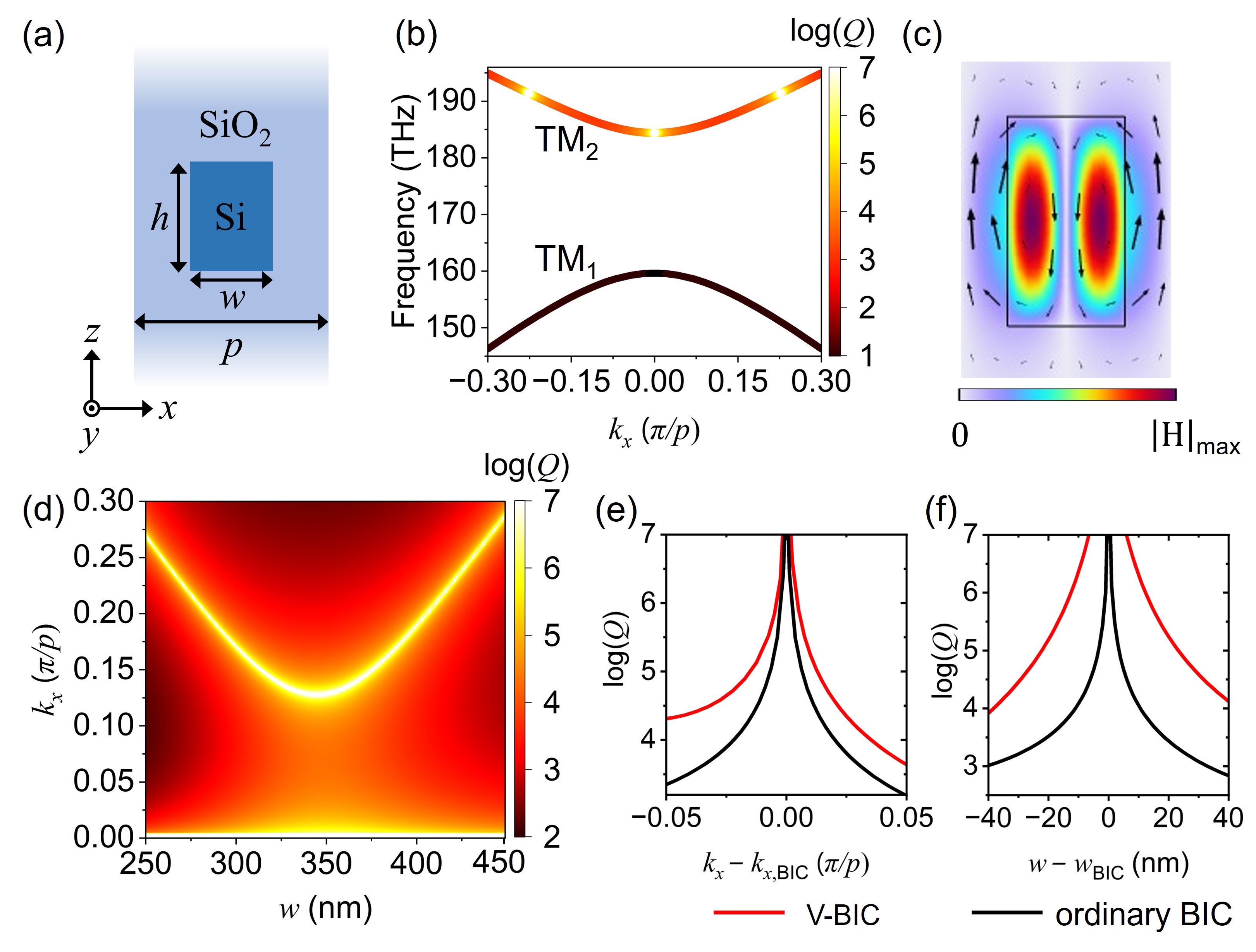}
    \caption{Example of a V-BIC in a 1D periodic grating. (a) Illustration of the grating geometry. The height of the grating bars \textit{h} is fixed at 750 nm in all the cases presented in this letter. (b) The band diagram of TM-polarized modes of the grating ($p$ = 760 nm, $w$ = 420 nm). The \textit{Q} factor is encoded in the  color. (c) The magnetic field distribution in one unit cell of the grating for the symmetry-protected BIC at the $\Gamma$-point in the TM$_2$ band. The grating bar is shown by the black contour, the color shows the normalized magnetic field distribution, and the black arrows show the electric field vectors. (d) The \textit{Q} factor of the TM$_2$ mode as a function of the in-plane momentum $k_x$ and grating bar width $w$ for $p$ = 760 nm. The BICs are clearly visible as the high-\textit{Q} regions, with the accidental BIC following a ``V''-shaped trajectory. (e) \& (f) The $Q$ factors of a V-BIC (red curve, $w_{\text{BIC}}$ = 345 nm, $k_{x,\text{BIC}}$ = 0.1275 $\pi /p$) and an ordinary BIC (black curve, $w_{\text{BIC}}$ = 440 nm, $k_{x,\text{BIC}}$ = 0.2688 $\pi /p$) as functions of (e) the in-plane momentum and (f) the grating bar width.}
\end{figure}

In this letter, we present another approach, that allows one to achieve BICs with an extreme $Q$ factor and high robustness. By tailoring the geometry of a one-dimensional (1D) periodic grating, we obtain accidental BICs and force them to follow a bent trajectory in the parameter space of the in-plane momentum and one of the geometric parameters. The trajectory resembles letter ``V'', and therefore, we refer to such a BIC as a ``V-BIC''. Next, we choose a set of parameters, for which the tip of the ``V''-shaped trajectory approaches the symmetry-protected BIC at the $\Gamma$-point. This gives rise to a ``K''-shaped high-$Q$ region in the parameter space. The middle part of this region contains a BIC exhibiting a $Q$ factor that is much more robust to variations of the in-plane momentum and the geometric parameters. We further demonstrate the advantages of such a ``K-BIC'' in finite structures, highlighting its significance to practical applications. We explain the underlying mechanism of the V-BIC trajectory on the basis of coupling between a guided resonance and a Fabry-Pérot (FP) mode~\cite{phasediagram}. The coupling leading to the formation of V- and K-BICs can be achieved by adjusting the geometric parameters of the structure, which makes it universally applicable to many different planar periodic structures. Our findings contribute to a deeper understanding of BICs and can be used to develop extremely robust ultrahigh-$Q$ optical resonators.

Consider a grating made of silicon (Si) bars embedded in silicon oxide (SiO$_2$). The unit cell of the structure is shown in Fig. 1(a). The optical constants of the two materials are based on Refs. \cite{index-si1,index-si2,index-sio2,polyanskiy24}. We use the eigenfrequency analysis in the COMSOL Multiphysics software to calculate the eigenmodes of the grating, setting the Floquet periodic boundary conditions along the $x$ direction. Figure 1(b) shows an example band structure of the transverse magnetic (TM) modes (comprising $H_y$, $E_x$, and $E_z$ components) of a grating with dimensions: $p$ = 760 nm, $h$ = 750 nm and $w$ = 420 nm defined in Fig. 1(a). The upper band (TM$_2$) supports a symmetry-protected BIC at the $\Gamma$-point and two off-$\Gamma$ accidental BICs, indicated by the maxima of $Q$ factor encoded in the color. The magnetic-field distribution of the symmetry-protected BIC in the unit cell is shown in Fig. 1(c). By gradually tuning one of the geometric parameters, the accidental BICs can be moved either towards or away from the symmetry-protected BIC. However, in the case of a V-BIC, the dependence on the in-plane momentum ($k_x$) and the grating bar width ($w$) can be made non-monotonic. As shown in Fig. 1(d), the trajectory of the V-BIC in the $w$-$k_x$ parameter space forms a rounded ``V'' shape. Increasing $w$ from 250 nm to 345 nm moves the accidental BIC towards the $\Gamma$-point, but as $w$ is further increased from 345 nm to 550 nm, the accidental BIC moves away from the $\Gamma$-point. Such a trajectory has a vanishing slope ($dk_x/dw$) at the turning point ($w$ = 345 nm), which leads to an enhanced robustness of the $Q$ factor with respect to both $k_x$ and $w$. Figures 1(e) and 1(f) compare the $k_x$ and $w$ dependence of the $Q$ factor for the V-BIC taken at the turning point (red curve) and an ordinary BIC taken far from the turning point (black curve). There is a clear enhancement of the $Q$ factor of the V-BIC over a wide range of $k_x$ and $w$ compared to that of the ordinary BIC.

By decreasing $p$, the V-BIC presented in Fig. 1(d) can be moved along $k_x$ to approach the symmetry-protected BIC, as shown in Fig. 2(a). Decreasing $p$ down to 742 nm gives rise to a ``K''-shaped distribution of high $Q$ values in the $k_x$-$w$ parameter space. We refer to the BIC in the middle of this distribution as the K-BIC. Figure 2(b) shows that further decreasing $p$ (down to $p$ = 650 nm) splits the K-BIC into two ordinary merging BICs, located at two separate points along the $w$-axis. 
\begin{figure}[hbt!]
    \centering
    \includegraphics[width=1\linewidth]{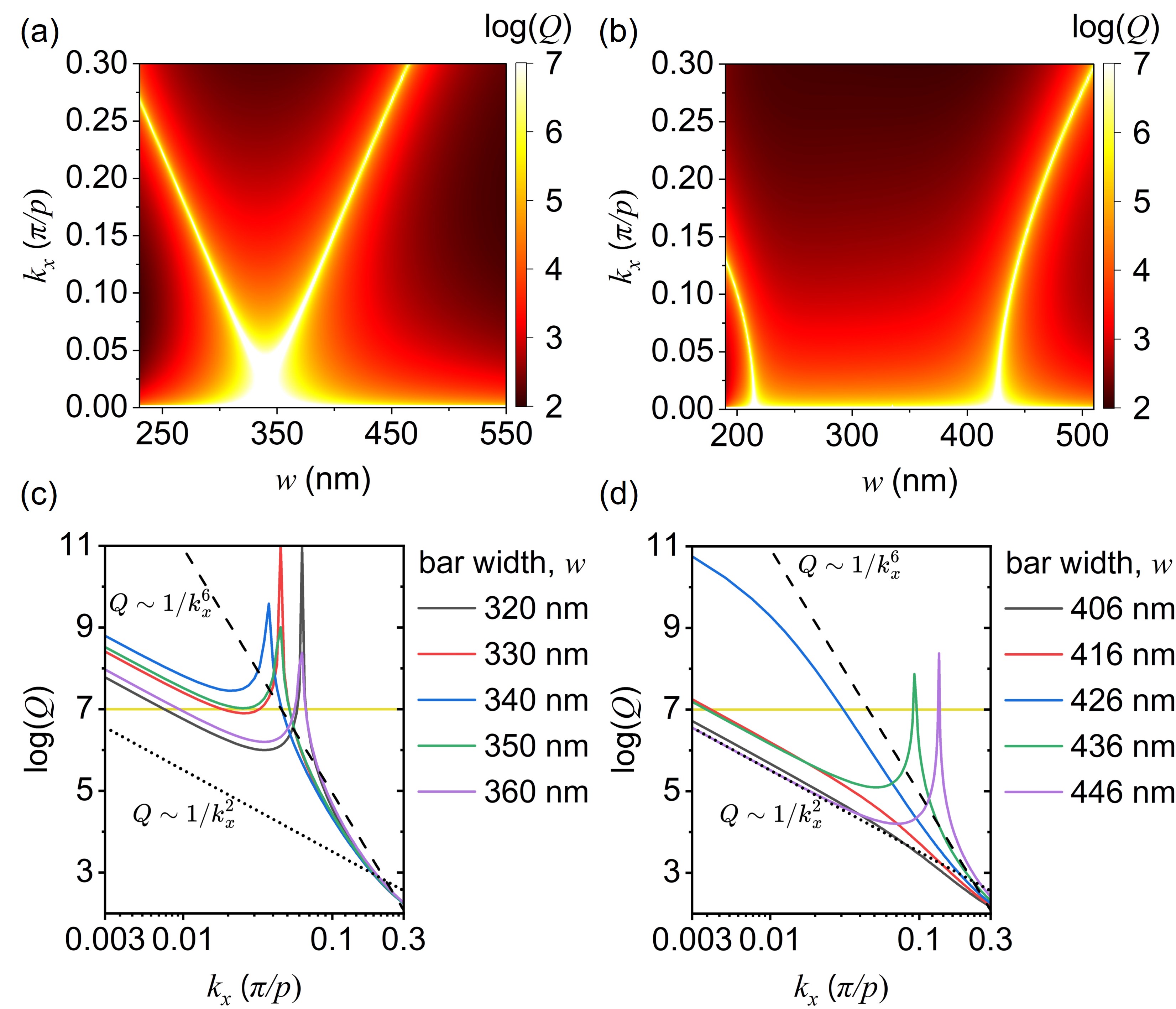}
    \caption{Example of an under-merged K-BIC and its comparison to ordinary merging BICs. (a) and (b) show the $Q$ factor as a function of $w$ and $k_x$ for the K-BIC (\textit{p} = 742 nm) and ordinary merging BICs (\textit{p} = 650 nm), respectively. (c) and (d) provide a comparison of the dependence of the $Q$ factor on $k_x$ and its robustness against deviations of $w$ for the K-BIC and an ordinary merged BIC, respectively. The scaling laws of $Q\sim1/k_x^6$ and $Q\sim1/k_x^2$ are shown by the dashed and dotted black lines, respectively. The solid yellow lines show the level of $Q=$ 10$^7$, which we consider as the practical upper limit for the experimentally achievable $Q$ factor.}
\end{figure}

Compared to the ordinary merging BICs, the K-BIC shows a significantly enhanced $Q$ factor over a much wider range of $k_x$ and $w$. This has important practical implications for the real-world structures, in which finite size and fabrication imperfections lead to discretization of the optical modes along $k_x$ and their mutual coupling~\cite{merging,laser-minibic}. The radiation channels of all the modes contribute to the overall loss of the excited resonances, depending on the coupling strength. Therefore, expanding the high-$Q$ range in the momentum space to reduce the radiative losses of all the nearby resonances is crucial for achieving higher $Q$ factors in practical samples. Furthermore, extending the high-$Q$ range of the geometric parameters increases the fabrication tolerance. 

In the example shown in Fig. 2(a), we have tailored the system such that the turning point of the V-BIC moves close to the symmetry-protected BIC, but does not touch it. The resulting BIC is therefore not a merged BIC, but the one in an under-merged state. Previously, it has been observed that such under-merged BICs can exhibit superior performance over the exactly merged BICs~\cite{bic-mini,laser-minibic}. We find that the K-BIC considered here provides the widest $k_x$ range of $Q>$ 10$^7$, which we consider as the practical upper limit for a BIC in any realistic structure (due to unavoidable scattering and absorption losses associated with available nanofabrication methods and finite size effects). Note that an exact mirror image of the distribution in Fig. 2(a) exists at negative values of $k_x$ due to the symmetry of the structure, which doubles the $k_x$-range of high $Q$ factors. While the $Q$ factor near the $\Gamma$ point could be further increased by tuning $p$ towards complete merging, this would decrease the $Q$ factor at larger $k_x$.
Figures 2(c) and 2(d) give a clear view on the enhanced robustness of the K-BIC compared to an ordinary merged BIC. Despite shifting $w$ off the optimal value by 20 nm, the K-BIC still retains a decently high $Q$ factor along a wide $k_x$ range, with a scaling law close to $Q\sim1/k_x^6$, while for the ordinary merging BIC, a 20 nm distortion would nearly eliminate the $Q$ factor enhancement and degrade the scaling law down to $Q\sim1/k_x^2$.

\begin{figure}[b]
    \hspace{0pt}
    \centering
    \includegraphics[width=1\linewidth]{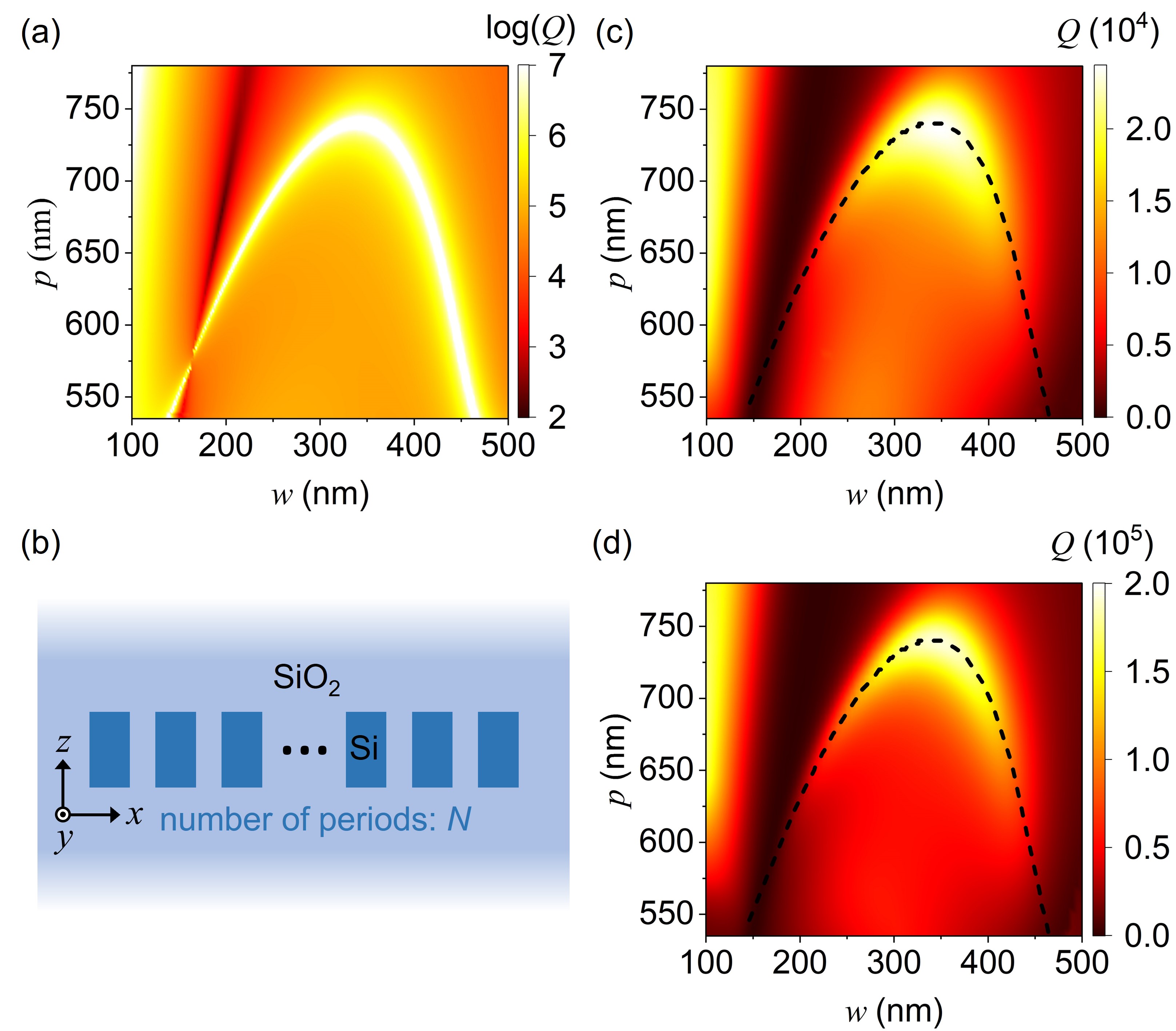}
    \caption{Enhanced $Q$ factor robustness due to K-BIC in both infinite and finite gratings. (a) The $Q$ factor distribution in the $w$-$p$ parameter space at $k_x$ = 0.01 $\pi /p$ for an infinite grating. The turning point of the trajectory corresponds to the K-BIC. (b) Illustration of a finite grating with $N$ periods. (c) \& (d) $Q$ factor distributions in the $w$-$p$ parameter space for finite gratings with $N$ = 50 and $N$ = 100 periods, respectively. The trajectory for the infinite grating obtained in (a) is plotted as the black dashed line.}
\end{figure}

The $Q$ factor diverges to infinity at the $\Gamma$ point regardless of the BIC type. Therefore, in order to investigate the boost of the $Q$ factor as a function of two geometric parameters, we probe the values of $Q$ at a fixed non-zero in-plane momentum close to the $\Gamma$-point. Such a dependence is shown in Fig. 3(a) for $k_x$ = 0.01 $\pi /p$, mapping the trajectory of the highest $Q$-factor boost at that point in the momentum space. The K-BIC is located approximately at the turning point of the trajectory. In addition, the dependence reveals a line of reduced $Q$ factor associated with a spectral overlap of the high-$Q$ mode with another mode of much lower $Q$.

From a practical perspective, one should demonstrate that the proposed K-BICs have superior properties in the case of finite gratings (Fig. 3(b)). Figures 3(c) and 3(d) show the $Q$ factor of the fundamental BIC mode in gratings consisting of 50 and 100 periods, respectively, as a function of $w$ and $p$ (similar to Fig. 3(a)). In both cases, the $Q$ factor is enhanced the most at the turning point of the trajectory (black dashed curve) corresponding to the K-BIC of the infinite grating. It can be seen that the $Q$ factor is also noticeably increased over an extremely wide region in the $w$-$p$ space around the turning point of the trajectory. These finding suggests that the K-BIC mechanism could be used to greatly improve the $Q$ factors of optical resonances and their robustness in finite periodic structures.

\begin{figure}
    \centering
    \includegraphics[width=1\linewidth]{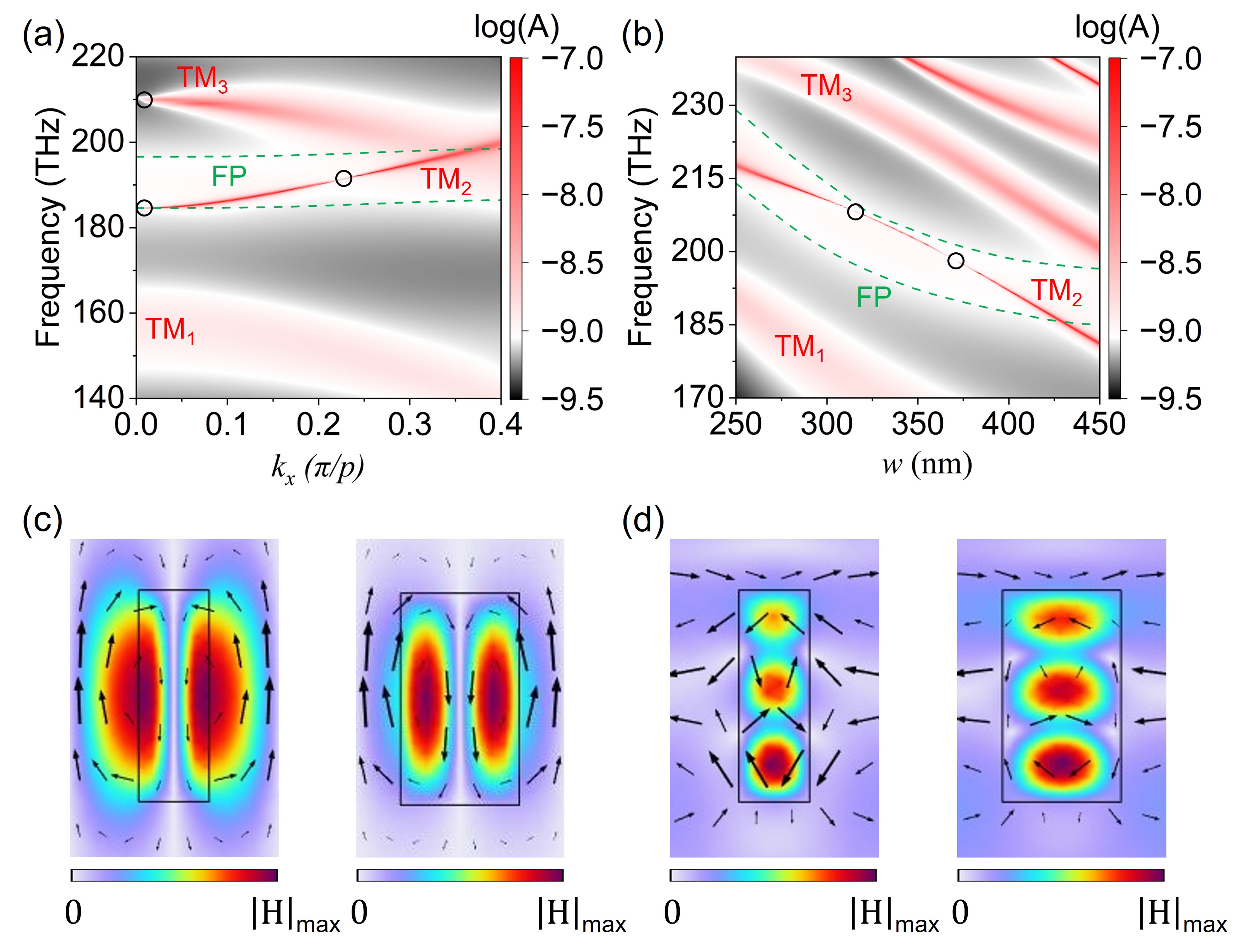}
    \caption{Revealing the underlying mechanism of the formation of V-BICs. (a) Absorption spectrum obtained with TM-polarized excitation ($p$ = 760 nm, $w$ = 420 nm), showing several modes including three TM guided resonances (red bands) and the FP mode (white band marked with green dashed lines). The BICs are indicated by the vanishing bandwidth and marked with circles. (b) Absorption spectrum obtained at a fixed incidence angle ($k_x$ = 0.14 $\pi /p$) while changing $w$ ($p$ = 760 nm), showing that the dispersion curves of the TM$_2$ and FP modes are bent towards opposite directions. (c) \& (d) Magnetic field distributions obtained from the scattering simulations of (c) the TM$_2$ mode and (d) the FP mode. In (c) and (d), the left panels are for $w$ = 250 nm and the right panels for $w$ = 420 nm. The black arrows show the electric field vectors.}
\end{figure}

In the remaining part of this letter, we show that the bent trajectories of V-BICs originate from distinct dispersion curves of the two constituent modes of the BICs: the TM$_2$ mode and a Fabry-Pérot mode. The FP mode exhibits a strong leakage to the far-field, making it inaccessible via the eigenfrequency analysis. Therefore, we perform scattering simulations to characterize its dispersion. We artificially introduce a tiny imaginary part to the refractive index of the grating material to reveal the optical modes through the increase of the absorbed optical power. The absorption spectrum obtained for TM-polarized incidence is presented in Fig. 4(a), effectively mapping the band structure, which shows also the modes TM$_1$ and TM$_2$ presented in Fig. 1(b). The BICs are marked by black circles. The accidental BICs belonging to the modes of the TM$_2$ band arise from coupling of the band to a FP mode~\cite{phasediagram} visible as a wide band marked by green dashed lines. Figure 4(b) shows the absorption spectrum of the structure as a function of $w$ under a fixed incidence angle (corresponding to $k_x$ = 0.14 $\pi /p$). The mode frequency ($\omega$) is seen to decrease when $w$ increases at a fixed period, which is explained by the fact that the effective refractive index of the modes ($1/n_{eff}$) increases together with the amount of silicon in the structure. Indeed, when the propagation constant is tied to the period, $\omega$ is approximately proportional to $1/n_{eff}$. However, the absorption curves of the TM$_2$ and FP modes have opposite concavities. When parameter $p$ changes, the curves move with respect to each other in the vertical direction and their two intersection points either approach each other or move further apart. The accidental BICs are associated with these intersection points, which explains the non-monotonic behavior of the dashed line in Fig. 3(a).

To understand the opposite concavities of the TM$_2$ and FP modes, we analyze their magnetic field distributions obtained from the scattering simulations (see Figs. 4(c-d)). When $w$ decreases, the power of the TM$_2$ mode is pushed out of the silicon strips into glass. As a result, at small values of $w$, the mode frequency becomes nearly constant. Since at increased $w$, the mode frequency must decrease, the curve in Fig. 4(b) is concave downward. In contrast, the power of the FP mode is almost completely confined in silicon, especially if $w$ is large, which makes the mode index essentially constant at large values of $w$. Hence, the FP curve is concave upward. To create a V-BIC, one needs to engineer the dispersion curves of the BIC's constituent modes to have opposite concavities in their dependence on one of the parameters. Building upon this generic method, future research could unlock the emergence of new families of robust BICs with unprecedented properties.

In conclusion, we have discovered BICs that exhibit a significantly improved $Q$ factor and its robustness in both finite and infinite periodic structures. The superior properties of these BICs result from their V- or K-shaped momentum-space trajectory that is a non-monotonic function of the geometric parameters. In particular, by tuning one of these parameters, we made the accidental BIC approach the $\Gamma$-point and then move away from it after reaching a turning point. Tailoring the system such that the turning point is located close to the $\Gamma$-point, we have created an under-merged BIC that exhibits a high $Q$ factor over a much wider in-plane momentum range compared to the ordinary merging BICs. We have shown that such a BIC can provide a significant increase of the $Q$ factor in finite gratings, additionally making it extremely robust to geometry distortions. We have explained the formation mechanism of the V-BICs on the basis of different dispersions of the guided resonance and the FP mode that couple together to give rise to the accidental BIC. The proposed mechanism can be extended to a wide range of photonic structures beyond simple 1D gratings. Our findings may enable ultrahigh-$Q$ optical resonators with exceptional robustness and excellent fabrication tolerances.

The authors acknowledge the support of the Academy of Finland (Grants No. 347449 and 353758). For computational resources, the authors acknowledge the Aalto University School of Science “Science-IT” project and CSC – IT Center for Science, Finland.

\bibliography{apssamp}% Produces the bibliography via BibTeX.
\end{document}